# Good Code Sets from Complementary Pairs via Discrete Frequency Chips

**Ravi Kadlimatti [1,]*and Adly T. Fam [2]**

[1] Advanced Wireless Systems Research Center, State University of New York at Oswego, Oswego, NY 13126, USA; ravi.kadlimatti@oswego.edu

[2] Department of Electrical Engineering, University at Buffalo, The State University of New York, Buffalo, NY 14260, USA; afam@buffalo.edu

* Correspondence: ravi.kadlimatti@oswego.edu; Tel.: +1-716-491-8985

**Abstract:** It is shown that replacing the sinusoidal chip in Golay complementary code pairs by special classes of waveforms that satisfy two conditions, symmetry/anti-symmetry and quazi-orthogonality in the convolution sense, renders the complementary codes immune to frequency selective fading and also allows for concatenating them in time using one frequency band/channel. This results in a zero-sidelobe region around the mainlobe and an adjacent region of small cross-correlation sidelobes. The symmetry/anti-symmetry property results in the zero-sidelobe region on either side of the mainlobe, while quasi-orthogonality of the two chips keeps the adjacent region of cross-correlations small. Such codes are constructed using discrete frequency-coding waveforms (DFCW) based on linear frequency modulation (LFM) and piecewise LFM (PLFM) waveforms as chips for the complementary code pair, as they satisfy both the symmetry/anti-symmetry and quasi-orthogonality conditions. It is also shown that changing the slopes/chirp rates of the DFCW waveforms (based on LFM and PLFM waveforms) used as chips with the same complementary code pair results in good code sets with a zero-sidelobe region. It is also shown that a second good code set with a zero-sidelobe region could be constructed from the mates of the complementary code pair, while using the same DFCW waveforms as their chips. The cross-correlation between the two sets is shown to contain a zero-sidelobe region and an adjacent region of small cross-correlation sidelobes. Thus, the two sets are quasi-orthogonal and could be combined to form a good code set with twice the number of codes without affecting their cross-correlation properties. Or a better good code set with the same number codes could be constructed by choosing the best candidates form the two sets. Such code sets find utility in multiple input-multiple output (MIMO) radar applications.

## 1. Introduction

*1.1. Background*

Good aperiodic codes are characterized by a narrow mainlobe and small sidelobes. Smaller autocorrelation peak sidelobes reduce the probability of false alarm, while a narrower mainlobe enhances the range resolution. Such properties are desirable in certain communications applications like preamble synchronization and in most radar applications.

Let $x[n]$ be a code of length $N$. Its aperiodic autocorrelation (ACF) can be represented by,

$$R[n] = \sum_{k=0}^{N-1} x[k]x^*[n+k] \quad (1)$$

In the $Z$-domain, if $x[n] \overset{Z}{\Leftrightarrow} X(z)$ then, its ACF can be represented by,

$$R[n] \overset{Z}{\Leftrightarrow} R(z) = X(z)X(z^{-1}) = N + S(z) \qquad (2)$$

where $N$ and $S(z)$ represent the mainlobe/peak and the sidelobes respectively.

Let $p(t) = e^{j2\pi f_p t}$, $0 \le t \le T$, represent a sinusoidal chip. $x[n]$ modulates $p(t)$ resulting in $x_p(t)$.

$$x_p(t) = \sum_{n=0}^{N-1} x[n]\, p(t - nT) \qquad (3)$$

Its ACF can be represented by,

$$R_x(t) = \int_{t}^{t+T} x_p(\tau) x_p^*(t+\tau) d\tau \qquad (4)$$

If $x_p(t)$ is sampled at intervals of $T_s = \frac{1}{f_s}$, then in the $Z$-domain, $x_p(mT_s)$ for $m = 0, 1, 2, \ldots, (N-1)N_s$, can be represented by,

$$x_p(mT_s) \overset{Z}{\Leftrightarrow} X_p(z) = X(z^{N_s})P(z) \qquad (5)$$

where $p(mT_s) \overset{Z}{\Leftrightarrow} P(z)$ and $N_s$ represents the number of samples in $P(z)$. ACF of $X_p(z)$ can be represented by,

$$R_x(z) = X_p(z)X_p(z^{-1}) \qquad (6)$$

$$R_x(z) = X(z^{N_s})P(z)X(z^{-N_s})P(z^{-1}) \qquad (7)$$

$$R_x(z) = R(z^{N_s})R_p(z) \qquad (8)$$

$$R_x(z) = NR_p(z) + S(z^{N_s})R_p(z) \qquad (9)$$

where $R_p(z) = P(z)P(z^{-1})$ represents the ACF of the highly sampled chip and $R(z)$ is given in Equation (2). It can be observed that the mainlobe and the sidelobes are spread by the ACF of the chip ($R_p(z)$).

Some of the well-known good aperiodic codes are the Barker codes, Frank codes [1] and waveforms based on Costas arrays [2–4]. Barker codes are biphase codes that have unity magnitude peak sidelobes. However, the longest known Barker code of odd length is of length 13 which gives a peak sidelobe ratio of $\frac{1}{13}$. Frank codes are polyphase codes obtained by concatenating the rows of a discrete Fourier transform (DFT) matrix. Thus, Frank codes are available for a variety of code lengths and achieve high peak sidelobe ratios for long code lengths. However, as the code length increases the number of distinct phases also increases. Costas arrays are completely classified to date up to order 27, and are known to exist of orders up to 200. They are widely used in radar and sonar applications because of their thumbtack ambiguity function. The codes presented in this paper are based on biphase complementary code pair that are available at a variety of code lengths.

One of the widely used approaches to suppress the ACF sidelobes of a waveform/code is via windowing whereby the peak sidelobe level is reduced, but at the cost of increasing the mainlobe width. Mismatched filters like the ones introduced in [5–8] could also be used to reduce ACF sidelobes at the cost of a small loss in signal-to-noise ratio. In [9], it is shown that complete sidelobe cancellation for a class of aperiodic codes is possible using additive-multiplicative processing of the matched filter (MF) output. However, the non-linear processing involved in these mismatched filters incurs additional computational cost and suffer from degraded performance in high noise environments. Although it is impossible for individual aperiodic codes to have zero sidelobes, Golay complementary code pair [10], polyphase complementary codes [11] and complementary code sequences [12] achieve complete sidelobe cancellation via addition of their autocorrelations. Figure 1 shows the transmission and reception of a Golay complementary code pair (say $a[n]$ and $b[n]$, of length $N_g$). $a[n]$ is transmitted at frequency $f_1$ (using $c_1(t) = e^{j2\pi f_1 t}$) and $b[n]$ at $f_2$ (using

$c_2(t) = e^{j2\pi f_2 t}$). At the receiver, the code pair are received in their own matched filters. Addition of their MF outputs ($R_a(t)$ and $R_b(t)$) results in complete cancellation of the sidelobes. However, frequency selective fading results in unequal attenuation of the two codes. This results in reemerging sidelobes due to inexact cancellation of the non-zero sidelobes of the code pair, as shown in Figure 2 for $N_g = 8$. Since the sidelobes of the code pair are not small, any inexact cancellation results in reemerging sidelobes which is highly undesirable. In Figure 2, $R_a(t)$, $R_b(t)$ and $R(t)$ represent the fading absent case and $R_a'(t)$, $R_b'(t)$ and $R'(t)$ represent the fading present case. For the fading present case, $a(t)$ and $b(t)$ are attenuated by 0.95 and 0.85 respectively. Galati et al. [13] also discusses in detail this effect of unequal attenuation on the complementary code pairs. Liu et al. [14] describes z-complementary code pairs that achieve a zero-correlation zone around the mainlobe and minimum possible sidelobe peaks outsize the zero domain. However, these codes also require two frequencies or channels which makes them vulnerable to the aforementioned effects of fading. Tang et al. [15] and Li [16] describe (loosely synchronized) LS and (large area) LA codes for (quasi-synchronous code division multiple access) QS-CDMA systems. These codes also achieve a zero interference zone in both the autocorrelation and the cross-correlations, but the peak sidelobes outside the zero-sidelobe region are not small. In this paper, we introduce codes constructed from biphase Golay complementary code pairs that are available at a variety of lengths. These codes have a zero-sidelobe domain on either side of the mainlobe, while the sidelobe peaks outside the zero domain are very small. Since these codes use the same frequency band, they are not affected by the effects of fading.

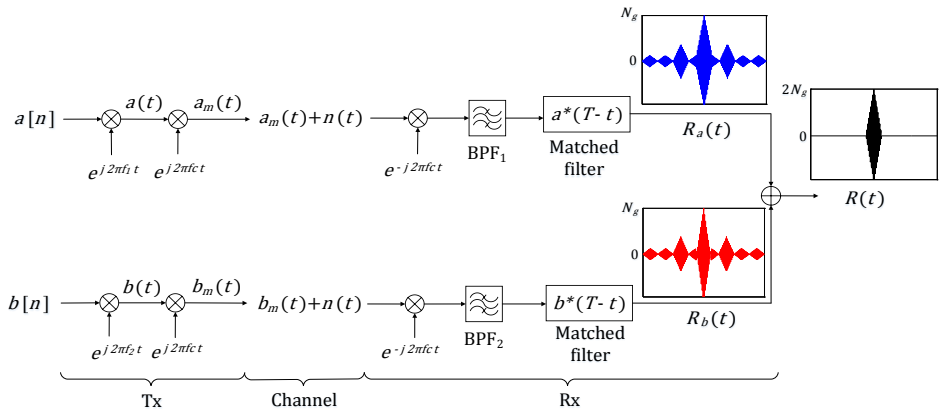

**Figure 1.** Transmitter (Tx) and receiver (Rx) of a Golay complementary code pair.

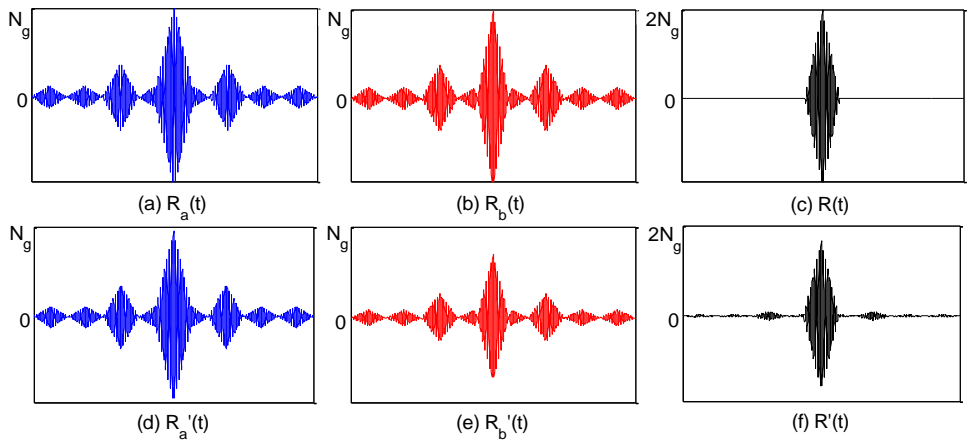

**Figure 2.** (**a**–**c**) Plots of matched filter outputs in the absence of frequency selective fading; and (**d**–**f**) plots of matched filter outputs in the presence of fading.

*1.2. The Proposed Codes Based on Complementary Pairs via Discrete Frequency Chips*

Given any complementary code pair $a[n]$ and $b[n]$, $a[n] + b[n - N_D]$ is constructed by concatenating $a[n]$ and $b[n]$ in the time domain using one frequency band/channel. The chips used for $a[n]$ and $b[n]$ satisfy the following two conditions:

1. They are symmetrical/anti-symmetrical mirror images of each other, i.e., $c(t)$ with $a[n]$ and $c(-t)/-c(-t)$ (or $\gamma c(-t)$, where $\gamma = 1$ or $-1$) with $b[n]$.
2. They are quasi-orthogonal in the convolution sense.

This concatenated code has a region of zero-sidelobes on either side of the mainlobe and an adjacent region of small cross-correlations. Due to the symmetry/anti-symmetry property, the two chips have identical ACF. This results in exact sidelobe cancellation of the complementary code pair, creating a zero-sidelobe region on either side of the mainlobe. The cross-correlation peak ($CCP$) between $c(t)$ and $\gamma c(-t)$ is represented by,

$$CCP = \max_t \left| \frac{\int_t^{t+T} \gamma c(\tau) c^*(-t + \tau) d\tau}{2 N_g} \right| \tag{10}$$

Since $c(t)$ and $\gamma c(-t)$ are quazi-orthogonal in the convolution, $CCP$ is small compared to the mainlobe peak $2N_g$. This property makes the adjacent region of cross-correlation sidelobes small.

Discrete frequency-coding waveforms (DFCW) based on linear frequency modulation (LFM, also known as a chirp signal) and piecewise LFM (PLFM) [17] waveforms satisfy both the symmetry/anti-symmetry and quasi-orthogonality conditions. Hence, they are used as chips for the complementary code pair. A discrete frequency-coding waveform, say $c(t)$, is the sum of $N$ contiguous sub-pulses of the same duration $\Delta T$, but not necessarily the same frequency.

$$c(t) = \frac{1}{\sqrt{\Delta T}} \sum_{n=0}^{N-1} e^{j2\pi f[n]\Delta W t_n} \tag{11}$$

where $\Delta W$ is the smallest possible frequency offset between two frequencies, $f[n]\Delta W$ is the frequency of the *n*th sub-pulse and $f[n] \in \{0, 1, 2, \ldots, N-1\}$.

DFCW sets find utility in MIMO radars [18] and Orthogonal Netted Radar Systems (ONRS) [19,20] that improve radar performance through spatial diversity. Several good DFCW sets have been proposed, such as the ones in [20–22].

In this paper, it is also shown that a quasi-orthogonal set of symmetrical/anti-symmetrical DFCW waveforms could be used as chips with the same complementary code pair to result in a good code set with a zero-sidelobe region. Several good code sets are constructed by changing the slopes/chirp rates of the DFCW chips based on LFM and PLFM waveforms. The code sets constructed with DFCW chips based on LFM waveforms occupy different bandwidths but have smaller peak cross-correlations, while the code sets constructed with DFCW chips based on PLFM waveforms occupy the same bandwidth but have slightly larger peak cross-correlations.

*1.3. Construction of a Good Code Set from the Mates of the Compelemntary Code Pair Resulting in Doubling the Number of Codes in the Set or a Better Good Code Set with Significaltly Smaller Cross-Correlations*

Given a complementary code pair, $a[n]$ and $b[n]$, there exists the complementary code pair: $b[-n + N_g]$ and $-a[-n + N_g]$, such that sum of the cross-correlation of $a[n]$ with $b[-n + N_g]$ and that of $b[n]$ with $-a[-n + N_g]$ results in complete sidelobe cancellation. The code pair: $b[-n + N_g]$ and $-a[-n + N_g]$, are called as the mates [23,24] of the complementary code pair: $a[n]$ and $b[n]$. Thus, two good code sets could be constructed from the complementary code pair and their mates using the same DFCW waveforms as chips. Since these two good code sets will be quasi-orthogonal to each other, they could be combined to form a larger good code set with double the number of codes in the set without affecting the cross-correlation properties. Instead of using all the codes in the two sets, it is shown that a better good code set with significantly reduced cross-correlation peaks could be constructed by choosing the best candidates form the two sets.

*1.4. Previous Research*

In [25,26], continuous frequency LFM and PLFM waveforms were used as chips for $a[n]$ and $b[n]$. The current manuscript uses the same code structure as in [25,26], but DFCW waveforms based on LFM and PLFM waveforms are used as chips for the complementary code pair. These codes find utility in more modern digital radar systems like the ones discussed in [17–19]. In addition, this paper also introduces a method of doubling the number of codes in the set without affecting the cross-correlation properties or constructing a better good code set with significantly smaller cross-correlations than the ones introduced in [25,26]. This is achieved by constructing good code sets from a complementary code pair and their mates, while using the same DFCW waveforms as chips. This is explained in detail in Section 7.

*1.5. Paper Structure*

Section 2 explains the two properties of symmetry/anti-symmetry and quasi-orthogonality for the discrete frequency-coding waveform chips based on LFM and PFLM waveforms. Construction of the proposed codes using DFCW chips is explained in Section 3. Doppler properties of the proposed codes are discussed in Section 4. Invariance of the zero-sidelobe region under frequency selective fading is demonstrated in Section 5. Section 6 shows the construction of good code sets. Section 7 shows the method of doubling the number of codes in the set or constructing a better good code set followed by conclusion in Section 8.

## 2. Symmetrical/Anti-Symmetrical DFCW Chips That Are Quasi-Orthogonal

Let $a(t)$ and $b(t)$ be constructed by using $c(t)$, for $0 \leq t \leq T$, and $\gamma c(-t)$ (where $\gamma = 1$ or $-1$) as chips for the Golay complementary code pair $\{a[n], b[n]\}$ of length $N_g$, respectively. $a(t)$ and $b(t)$ can be represented as shown in Equation (3).

If $a[n] \overset{Z}{\Leftrightarrow} A(z)$ and $b[n] \overset{Z}{\Leftrightarrow} B(z)$, then

$$a(mT_s) \overset{Z}{\Leftrightarrow} A_c(z) = A(z^{N_c})C(z) \tag{12}$$

$$b(mT_s) \overset{Z}{\Leftrightarrow} B_c(z) = \gamma B(z^{N_c})C(z^{-1}) \tag{13}$$

where $a(mT_s)$ and $b(mT_s)$ represent the sampled versions of $a(t)$ and $b(t)$, respectively, $N_c$ is the number of samples in $C(mT_s) \overset{Z}{\Leftrightarrow} C(z)$ and $m \in \{0, \pm 1, \pm 2, \ldots, \pm \infty\}$.

$A_c(z) + z^{-D}B_c(z)$ represents a code constructed by concatenating $A_c(z)$ and $B_c(z)$ with a gap $D$ in between. At the receiver, this code is passed into two matched filters: $A_c(z^{-1})$ which is the matched filter of $A_c(z)$, and $B_c(z^{-1})$ which is the MF of $B_c(z)$. The cross-correlation of $A_c(z) + z^{-D}B_c(z)$ with $A_c(z^{-1})$ results in,

$$R_{A_c}(z) = \left(A_c(z) + z^{-D}B_c(z)\right)A_c(z^{-1}) \tag{14}$$

$$R_{A_c}(z) = A(z^{N_c})C(z)A(z^{-N_c})C(z^{-1}) + \gamma z^{-D}B(z^{N_c})C(z^{-1})A(z^{-N_c})C(z^{-1}) \tag{15}$$

$$R_{A_c}(z) = R_A(z^{N_c})R_c(z) + \gamma z^{-D}R_{B,A}(z^{N_c})C^2(z^{-1}) \tag{16}$$

where $R_A(z) = A(z)A(z^{-1})$ represents the autocorrelation of $a[n]$, $R_c(z) = C(z)C(z^{-1})$ represents the chip ACF ($C(z^{-1})$ is the MF of $C(z)$), $R_{B,A}(z) = B(z)A(z^{-1})$ represents the cross-correlation of $b[n]$ with $a[n]$ and $\gamma C^2(z^{-1})$ represents the cross-correlation of $\gamma C(z^{-1})$ with $C(z^{-1})$.

Similarly, the cross-correlation of $A_c(z) + z^{-D}B_c(z)$ with $B_c^{-1}(z)$ can be represented by,

$$R_{B_c}(z) = \gamma R_{A,B}(z^{N_c})C^2(z) + \gamma^2 z^{-D}R_B(z^{N_c})R_c(z) \tag{17}$$

where $R_{A,B}(z) = A(z)B(z^{-1})$ represents the cross-correlation of $a[n]$ with $b[n]$, $R_B(z) = B(z)B(z^{-1})$ represents the ACF of $b[n]$ and $\gamma C^2(z)$ is the cross-correlation of $C(z)$ with $\gamma C(z)$.

Delaying $R_{A_c}(z)$ by $D$ and adding it to $R_{B_c}(z)$ results in,

$$R_s(z) = z^{-D}R_{A_c}(z) + R_{B_c}(z) \tag{18}$$

Since $\gamma^2 = 1$ and $R_A(z) + R_B(z) = 2N_g$, $R_A(z^{N_c})R_c(z) + \gamma^2 R_B(z^{N_c})R_c(z) = 2N_g R_c(z)$, Equation (17) becomes

$$R_s(z) = \gamma R_{A,B}(z^{N_c})C^2(z) + 2N_g z^{-D}R_c(z) + \gamma z^{-2D}R_{B,A}(z^{N_c})C^2(z^{-1}) \tag{19}$$

$R_s(z)$ contains the mainlobe ($2N_g R_c(z)$) that is spread by the chip ACF (i.e., $R_c(z)$), a zero-sidelobe region on either side of the mainlobe and an adjacent region of cross-correlation sidelobes (i.e., the first and the last terms in Equation (19)). If $\gamma = 1$, then $C(z)$ and $\gamma C(z^{-1}) = C(z^{-1})$ represent symmetrical waveforms. If $\gamma = -1$, then $C(z)$ and $\gamma C(z^{-1}) = -C(z^{-1})$ represent anti-symmetrical waveforms. For both the symmetry ($\gamma = 1$) and anti-symmetry ($\gamma = -1$) conditions, ACF of $C(z)$, i.e., $R_c(z) = C(z)C(z^{-1})$ is identical to the ACF of $\gamma C(z)$, i.e., $R_{\gamma c}(z) = \gamma^2 C(z^{-1})C(z) = R_c(z)$. As a result, exact sidelobe cancellation of the complementary code pair is achieved resulting in the zero-sidelobe region on either side of the mainlobe as shown in Equation (19). If $C(z)$ and $\gamma C(z^{-1})$ are also quasi-orthogonal in the convolution sense, i.e., their cross-correlation peak (as defined in Equation (10)) is small compared to the mainlobe peak ($2N_g$), then the first and the last terms in Equation (19) are small. It should be noted that using a sinusoidal chip for the complementary code pair will also result in a zero-sidelobe region around the mainlobe as it satisfies the symmetry/anti-symmetry condition. However, a sinusoidal chip is not quasi-orthogonal to its symmetrical/anti-symmetrical mirror image. Thus, the proposed codes with sinusoidal chip achieves zero-sidelobe region but the cross-correlation sidelobes outside the zero domain are large which is undesirable. Hence, any waveform that satisfies the two conditions, symmetry/anti-symmetry and quasi-orthogonality, could be used as chips for the complementary code pair in the proposed code design.

DFCWs based on LFM and PLFM waveforms are constructed by discretizing their instantaneous frequencies. These will be referred to as Discrete Frequency LFM (DF-LFM) and Discrete Frequency PLFM (DF-PLFM) waveforms, respectively. The DF-LFM and DF-PLFM waveforms satisfy the symmetry/anti-symmetry and the quasi-orthogonality conditions required for the chips in the proposed code design. The following three types of DFCWs are used as chips for the complementary code pair in the proposed code design:

1. DF-LFM waveforms.
2. DF-PLFM waveforms comprised of two up-chirp segments.
3. DF-PLFM waveforms comprised of an up-chirp followed by a down-chirp as in [21].

Let $u(t)$ and $d(t) = u^*(-t)$ represent a DF-LFM/DF-PLFM waveform and its symmetrical mirror image respectively. The DF-LFM waveforms can be represented by,

$$u(t) = \frac{1}{\sqrt{\Delta T}} \sum_{n=0}^{N-1} e^{j2\pi f[n]\Delta W t_n} \tag{20}$$

$$d(t) = \frac{1}{\sqrt{\Delta T}} \sum_{n=0}^{N-1} e^{j2\pi f[N-1-n]\Delta W t_n} \tag{21}$$

where $f[n] = n$, $\Delta W = \frac{1}{\Delta T}$, $\Delta T = \frac{T}{N}$, $n\Delta T \leq t_n \leq (n+1)\Delta T$ and $T$ is the time-duration of the DFCW chip.

The DF-PLFM waveforms comprised of two up-chirp segments can be represented by,

$$u(t) = \begin{cases} \dfrac{1}{\sqrt{\Delta T}} \displaystyle\sum_{n=0}^{\frac{N}{2}-1} e^{j2\pi f_1[n]\Delta W t_n} \\ \dfrac{1}{\sqrt{\Delta T}} \displaystyle\sum_{n=\frac{N}{2}}^{N-1} e^{j2\pi f_2[N-1-n]\Delta W t_n} \end{cases} \tag{22}$$

$$d(t) = \begin{cases} \dfrac{1}{\sqrt{\Delta T}} \displaystyle\sum_{n=0}^{\frac{N}{2}-1} e^{j2\pi f_2[n]\Delta W t_n} \\ \dfrac{1}{\sqrt{\Delta T}} \displaystyle\sum_{n=\frac{N}{2}}^{N-1} e^{j2\pi f_1[N-1-n]\Delta W t_n} \end{cases} \quad (23)$$

where $f_1[n] = round\left(\dfrac{2k(N-1)n}{N}\right)$ for $n = 0, 1, \ldots, \dfrac{N}{2} - 1$, $0 < k \leq 0.5$ and $f_2[n] = round\left((N-1)(1 - \dfrac{(1-k)2n}{N-1})\right)$ for $n = \dfrac{N}{2}, \dfrac{N}{2} + 1, \ldots, N - 1$.

The DF-PLFM waveforms comprised of an up-chirp and a down-chirp segment can be represented by,

$$u(t) = \begin{cases} \dfrac{1}{\sqrt{\Delta T}} \displaystyle\sum_{n=0}^{\frac{N}{2}-1} e^{j2\pi f_1[n]\Delta W t_n} \\ \dfrac{1}{\sqrt{\Delta T}} \displaystyle\sum_{n=\frac{N}{2}}^{N-1} e^{j2\pi f_2[n]\Delta W t_n} \end{cases} \quad (24)$$

$$d(t) = \begin{cases} \dfrac{1}{\sqrt{\Delta T}} \displaystyle\sum_{n=0}^{\frac{N}{2}-1} e^{j2\pi f_2\left[\frac{N}{2}-1-n\right]\Delta W t_n} \\ \dfrac{1}{\sqrt{\Delta T}} \displaystyle\sum_{n=\frac{N}{2}}^{N-1} e^{j2\pi f_1[N-1-n]\Delta W t_n} \end{cases} \quad (25)$$

Figure 3. shows the plots of instantaneous frequencies of the aforementioned DFCW chips. Let $f_u(t)$ and $f_d(t)$ represent the instantaneous frequencies of $u(t)$ and $d(t)$, respectively. $R_u(t)$, $R_d(t)$ and $R_{u,d}(t)$ represent the ACF of $u(t)$, ACF of $d(t)$ and cross-correlation between $u(t)$ and $d(t)$ respectively. They can be represented as shown in Equation (4). Since $u(t)$ and $d(t)$ are symmetrical mirror images of each other, their ACFs are identical, i.e., $R_u(t) = R_d(t)$, as shown in the plots in Figure 4. Since $u(t)$ and $d(t)$ are also quasi-orthogonal, $\max_t |R_{u,d}(t)|$ is very small as shown in their plots in Figure 4. For these plots, $k = 0.24$ and $N = 32$. $k$ closes to 0 results in a DF-LFM/DF-PLFM waveform that is close to a pure sinusoid, while $k$ close to 0.5 results in a very sharp slope.

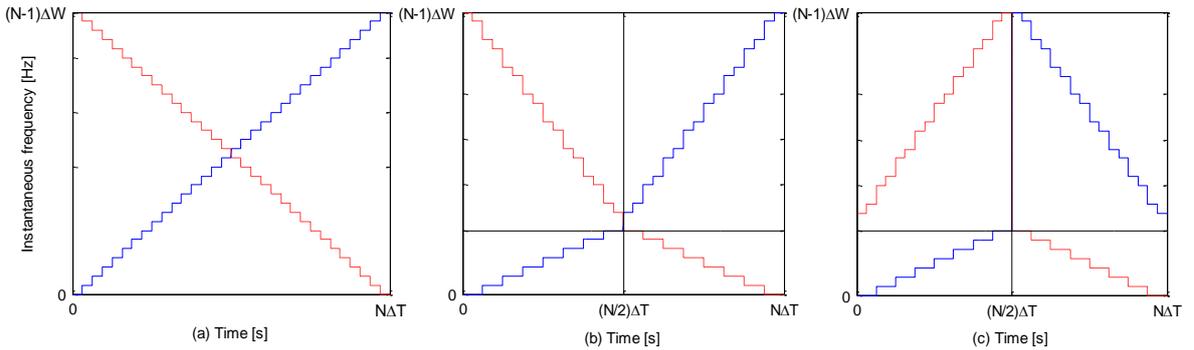

**Figure 3.** Plots of instantaneous frequencies of (a) discrete frequency linear frequency modulation (DF-LFM) chips, (b) discrete frequency piecewise linear frequency modulation (DF-PLFM) chips comprised of two up-chirp segments and (c) DF-PLFM chips comprised of an up-chirp and a down-chirp segment.

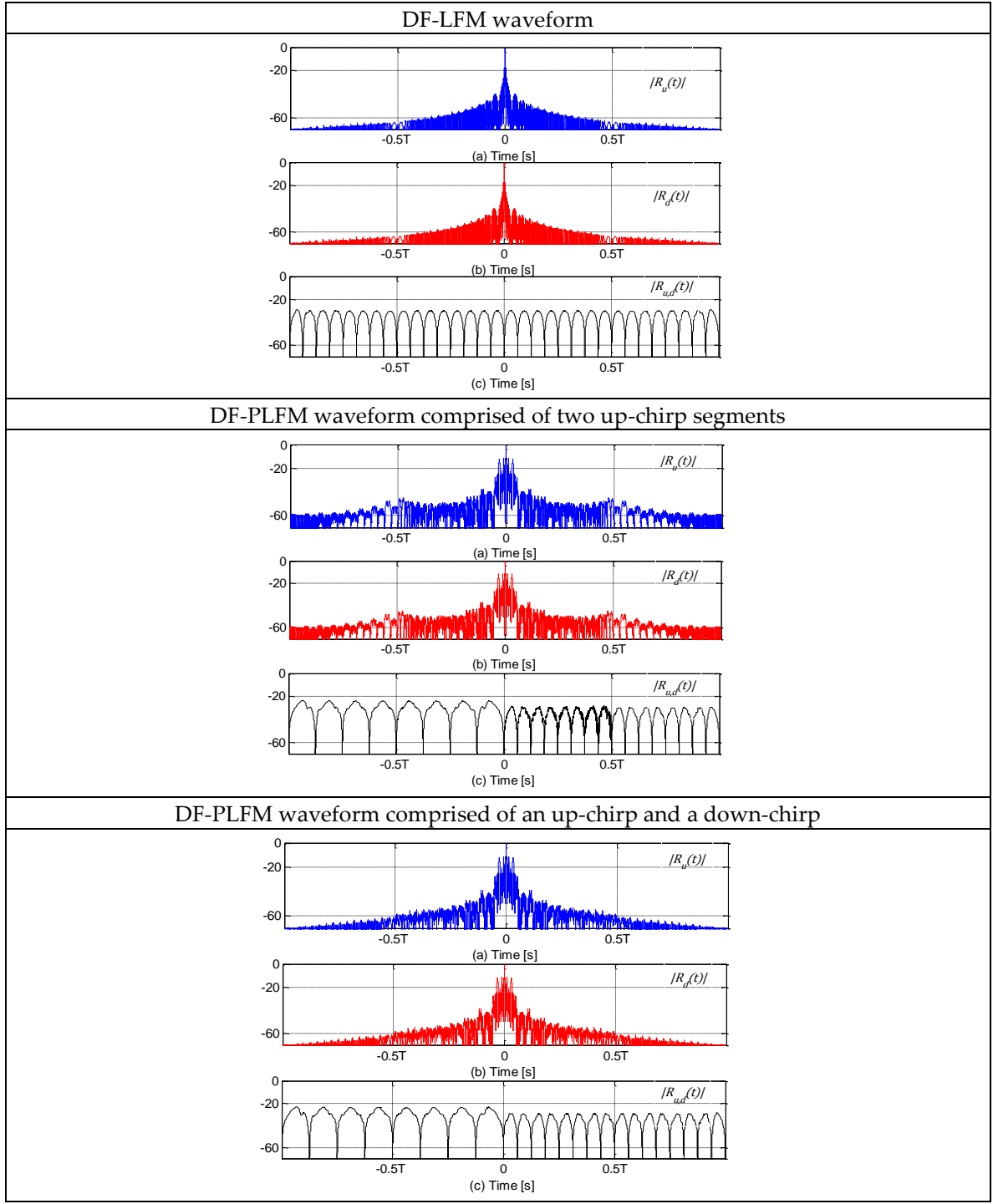

**Figure 4.** (a) and (b) Autocorrelation and (c) cross-correlation of the discrete frequency linear frequency modulation (DF-LFM) and discrete frequency piecewise linear frequency modulation (DF-PLFM) waveforms.

## 3. The Proposed Code Based on Complementary Pair Using DF-LFM and DF-PLFM Waveforms as Chips

Let $a[n]$ and $b[n]$ represent a Golay complementary code pair of length $N_g$. Let $a(t)$ and $b(t)$ represent the waveforms obtained by modulating $u(t)$ (Equation (20), (22) or (24)) with $a[n]$ and $d(t)$ (Equation (21), (23) or (25)) with $b[n]$, respectively.

$$a(t) = \sum_{n=1}^{N_g} a[n]u(t-(n-1)T) \qquad (26)$$

$$b(t) = \sum_{n=1}^{N_g} b[n]d(t-(n-1)T) \tag{20}$$

where $T = N\Delta T$ is the duration of $u(t)$ and $d(t)$.

A new code, $s(t) = a(t) + b(t - 2N_gT)$, is constructed by concatenating $a(t)$ and $b(t)$ with a delay of $2N_gT$ between them, as shown in Figure 5. $s(t)$ is then carrier ($f_c$) modulated before transmission, resulting in $s_m(t)$, as shown in Figure 5. The transmitted signal can be represented by,

$$s_m(t) = \left(a(t) + b(t - 2N_gT)\right)e^{j2\pi f_c t} \tag{21}$$

Figure 6 shows the block diagram of the receiver. The received signal, $x(t) = s_m(t) + n(t)$, is first carrier demodulated and then bandpass filtered (BPF), giving $x'(t)$ as shown in Figure 6. Since both $a(t)$ and $b(t)$ occupy the same bandwidth, only one BPF is used. $n \sim N(0, \sigma^2)$ represents additive white Gaussian noise.

$$x'(t) = a(t) + b(t - 2N_gT) + +n(t)e^{-j2\pi f_c t} \tag{22}$$

In one path, $x'(t)$ is passed into the matched filter (MF) of $a(t)$ implemented as a cascade of two MFs, one for the digital code ($a[n]$) and the other for the chip ($u(t)$), as shown in Figure 6.

$$R_{x',a}(t) = R_a(t) + R_{b,a}(t - 2N_gT) + n_a(t) \tag{30}$$

where $R_a(t)$ represents the ACF of $a(t)$, $R_{b,a}(t)$ represents the cross-correlation of $b(t)$ with $a(t)$ and $n_a(t)$ represents the filtered noise process, $n_a \sim N(0, N_g\sigma^2)$.

$$R_a(t) = \sum_{n=1}^{N_g}\sum_{m=1}^{N_g} a[n]a^*[N_g - m] \int_t^{t+T} u(\tau - (n-1)T)u^*(t + \tau + (m-1)T)d\tau \tag{31}$$

$$R_{b,a}(t) = \sum_{n=1}^{N_g}\sum_{m=1}^{N_g} b[n]a^*[N_g - m] \int_t^{t+T} d(\tau - (n-1)T)u^*(t + \tau + (m-1)T)d\tau \tag{32}$$

In another path, $x'(t)$ is passed into the MF of $b(t)$ which is also implemented as a cascade of MF of $b[n]$ and that of $d(t)$.

$$R_{x',b}(t) = R_{a,b}(t) + R_b(t - 2N_gT) + n_b(t) \tag{33}$$

where $R_{a,b}(t)$ represents the cross-correlation of $a(t)$ with $b(t)$, which can be represented as shown in Equation (17), $R_b(t)$ represents the ACF of $b(t)$ and $n_a(t)$ represents the filtered noise process, $n_b \sim N(0, N_g\sigma^2)$.

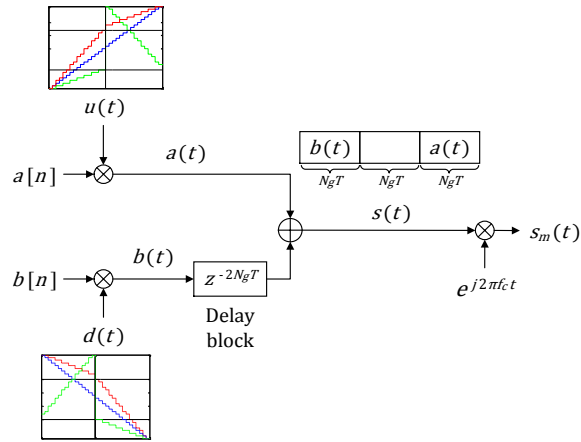

**Figure 5.** Block diagram showing the construction of the proposed code using Golay complementary code pair with DFCW chips.

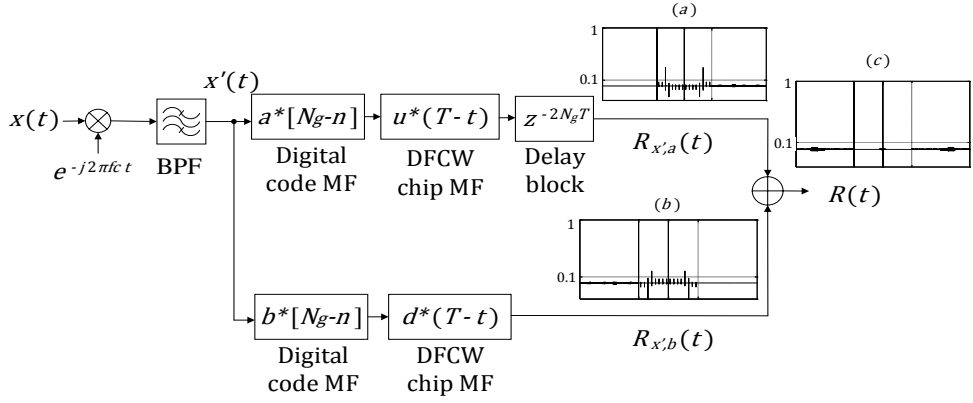

**Figure 6.** Receiver block diagram.

Delaying $R_{x',a}(t)$ by $2N_gT$ (as shown in Figure 6) and adding it to $R_{x',b}(t)$ results in,

$$R(t) = R_{a,b}(t) + R_a(t - 2N_gT) + R_b(t - 2N_gT) + R_{b,a}(t - 4N_gT) + n_a(t - 2N_gT) + n_b(t) \qquad (34)$$

Since,

$$R_a(t) + R_b(t) = \begin{cases} 2N_g R_u(t), -T \leq t \leq T \\ 0, otherwise \end{cases} \qquad (35)$$

where $R_u(t) = R_d(t)$ represents the ACF of the DFCW chip.

$$R(t) = R_{a,b}(t) + 2N_g R_u(t - 2N_gT) + R_{b,a}(t - 4N_gT) + n_a(t - 2N_gT) + n_b(t) \qquad (36)$$

Thus, the final output, $R(t)$, (excluding the noise terms) consists of,

$$R(t) = \begin{cases} R_{a,b}(t), 0 \leq t < 2N_gT \\ 0, 2N_gT \leq t < 3N_gT - T \\ 2N_g R_u(t), 3N_gT - T \leq t < 3N_gT + T \\ 0, 3N_gT + T \leq t < 4N_gT \\ R_{b,a}(t), 4N_gT \leq t < 6N_gT \end{cases} \qquad (37)$$

In Equation (37), $2N_g R_u(t)$ is the mainlobe of $R(t)$ ($R_u(t)$ is the ACF of the DFCW chip, as shown in the plots in Figure 4). $2N_gT \leq t \leq 3N_gT - T$ and $3N_gT + T \leq t \leq 4N_gT$ are the zero-sidelobe regions on either side of the mainlobe. $R_{a,b}(t)$ and $R_{b,a}(t)$ (defined in Equation (32)), respectively, represent the small cross-correlation sidelobes. Let $CCP$ represent the peak cross-correlation sidelobe. Since, $R_{a,b}(t)$ and $R_{b,a}(t)$ are mirror images of each other,

$$CCP = \max_t \left| \frac{R_{a,b}(t)}{2N_g} \right| \qquad (38)$$

The table in Figure 7 shows the plots of $|R_{x',a}(t - 2N_gT)|$, $|R_{x',b}(t)|$ and $|R(t)|$, all normalized by $2N_g$ and in the absence of noise, for $N_g = 8$ and $N = 16$ for the three DFCW chips in Column 1. 3D-plots of the corresponding cross-correlation peaks ($CCP$) for a range of $N_g$ and $N$ are shown in Column 2. Clearly, the ACF plots show the zero-sidelobe region around the mainlobe and the adjacent region of small cross-correlation sidelobes. From the 3D-plots of $CCP$ it can be observed that $CCP$ values are very small, and decrease in magnitude as the code length $N_g$ and the number of discrete frequencies $N$ in the DF-LFM/DF-PLFM chips increase. In addition, the $CCP$ values are slightly larger for the codes constructed using the DFCW chips based on PLFM waveforms compared to the $CCP$ values for the codes constructed using DFCW chips based on LFM waveform, but still very small ($\approx -30$ dB).

If $-a(t)$ and $b(t)$ are concatenated and transmitted at a second frequency band and received using a receiver similar to the one described in Figure 6, then the final output of this code can be represented by,

$$R'(t) = \begin{cases} -R_{a,b}(t), 0 \leq t < 2N_gT \\ 0, 2N_gT \leq t < 3N_gT - T \\ 2N_gR_u(t), 3N_gT - T \leq t < 3N_gT + T \\ 0, 3N_gT + T \leq t < 4N_gT \\ -R_{b,a}(t), 4N_gT \leq t < 6N_gT \end{cases} \qquad (39)$$

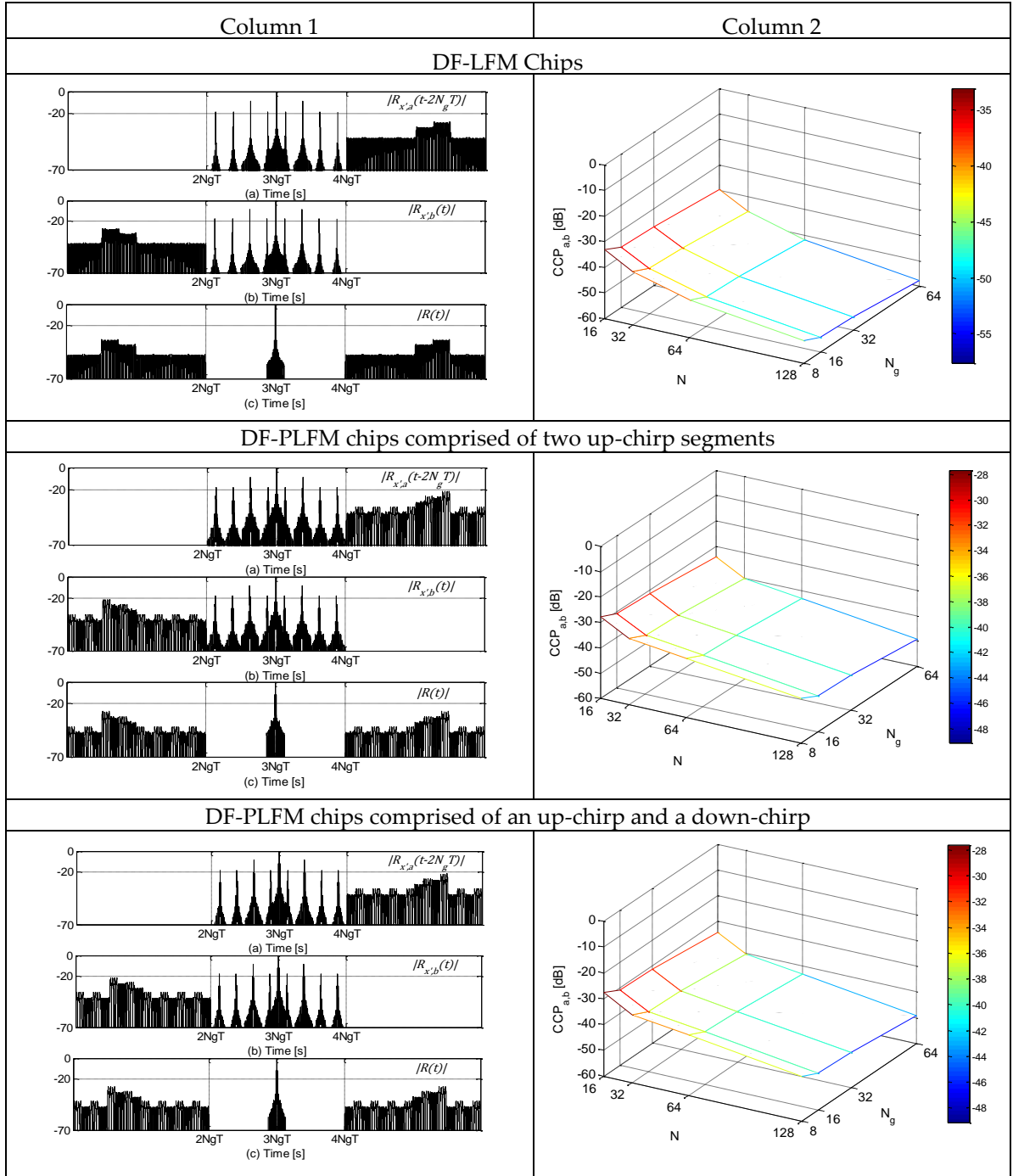

**Figure 7.** Column 1. (a), (b) and (c) Plots in dB of the normalized matched filter (MF) outputs. Column 2. Plots of the corresponding cross-correlation sidelobe peaks ($CCP$) vs. code length ($N_g$) vs. number of discrete frequencies in the DF-LFM/DF-PLFM chips ($N$).

Clearly, adding Equations (37) and (39) results in complete cancellation of the cross-correlation sidelobes. Thus, achieving complete zero-sidelobes on either side of the mainlobe. This is shown in the plots of $R(t)$, $R'(t)$ and $R(t) + R'(t)$ for $N_g = 8$ and $N = 16$ in Figure 8. Frequency selective fading between the two bands could result in inexact cancellation of the cross-correlation sidelobes that are adjacent to the zero-sidelobe region. However, since the cross-correlation sidelobes are very small, any inexact cancellation due to fading will not create undesirable high magnitude sidelobes.

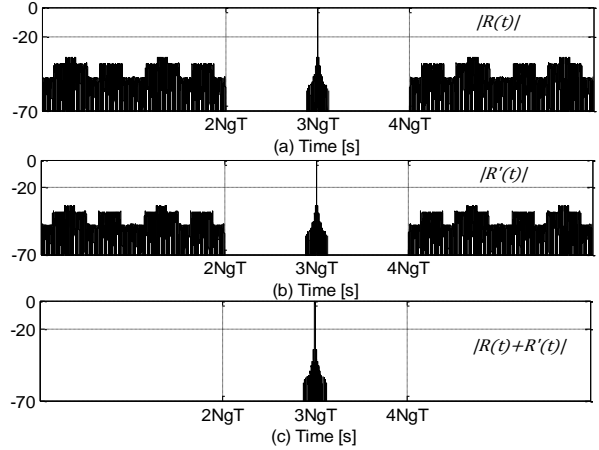

**Figure 8.** Plots in dB of: (**a**) $\frac{|R(t)|}{N_g}$; (**b**) $\frac{|R'(t)|}{N_g}$; and (**c**) $\frac{|R(t)+R'(t)|}{2N_g}$.

The complementary code pair could also be interlaced in the time domain (i.e., $a\left(\frac{t}{T}\right) + b\left(\frac{t}{T} - T\right)$). This scheme results in a transmission without gaps, while still using the same frequency band as in the previous scheme. Fishler et al. [18] and Deng [19] describe this scheme using continuous LFM and PLFM chips. However, for this interlaced code, the zero-sidelobes and the small cross-correlation sidelobes in the final output are interlaced as well.

## 4. Doppler Properties

Figure 9a,b shows the 3D ambiguity function (AF) of the proposed code constructed using DF-LFM chips for $N_g = 8$ and $N = 16$ as a function of the normalized Doppler frequency $f_d T$. The AF can be represented by,

$$R(t,f) = \int_{-\infty}^{\infty} s_f(\tau) s^*(t - \tau) d\tau \qquad (40)$$

where $s_f(t) = s(t)e^{j2\pi f_d t}$, for $0 \leq t \leq 3N_g T$, is the transmitted code Doppler shifted by $f_d$, $T$ is the duration of the DF-LFM/DF-PLFM chips and $s(t)$ is the proposed code.

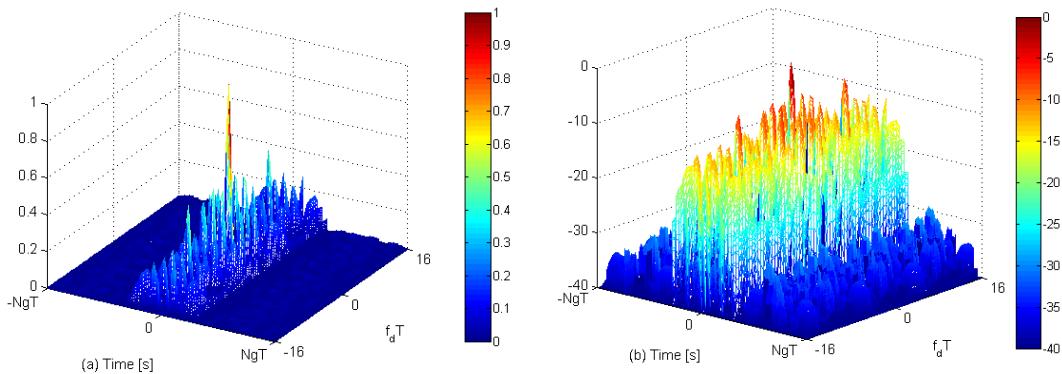

**Figure 9.** Plots of 3D-ambiguity function in (a) linear scale and (b) in dB scale of the proposed codes constructed using DF-LFM chips.

It can be observed that Doppler shift adversely affects the cancellation of the sidelobes of the complementary code pair, resulting in the disappearance of the zero-sidelobe domain on either side of the mainlobe. The ACF mainlobes of the symmetrical chips shift in opposite directions under

Doppler shift. As a result, ACF of the complementary code pair do not align with each other. Thus, the proposed codes do not have good Doppler properties. One way to improve the Doppler properties could be to use triangular FM based waveforms as chips for the complementary code pair in the proposed code structure. Another method to improve the Doppler properties could be to use the approach discussed in [27] to construct Doppler resilient Golay complementary code pairs.

## 5. Effect of Frequency Selective Fading

In the presence of frequency selective fading, the higher frequencies within the frequency sweep ($W = N(N-1)\Delta W \Delta T$) of the DF-LFM/DF-PLFM waveforms could be slightly attenuated compared to the lower frequencies. However, autocorrelations of the DF-LFM/DF-PLFM waveforms remain identical, since they are symmetrical. Hence, the zero-sidelobe region is not affected. However, the adjacent region of small cross-correlation sidelobes could vary slightly depending on the attenuation due to fading. However, its effect on the error in the location of the mainlobe is negligible as will be shown in the simulation results later in this section.

In the presence of frequency selective fading, let the lower and the higher frequency components of the DF-LFM/DF-PLFM chips be attenuated by $\alpha_1$ and $\alpha_2$ respectively, where $\alpha_2 < \alpha_1$. The DF-LFM chips ($u_f(t)$ and $d_f(t)$) in the presence of fading can be represented by,

$$u_f(t) = \begin{cases} \dfrac{\alpha_1}{\sqrt{\Delta T}} \sum_{n=0}^{N_1-1} e^{j2\pi f[n]\Delta W t_n} \\ \dfrac{\alpha_2}{\sqrt{\Delta T}} \sum_{n=N_1}^{N-1} e^{j2\pi f[n]\Delta W t_n} \end{cases} \tag{41}$$

$$d_f(t) = \begin{cases} \dfrac{\alpha_2}{\sqrt{\Delta T}} \sum_{n=0}^{N-1-N_1} e^{j2\pi f[N-1-N_1-n]\Delta W t_n} \\ \dfrac{\alpha_1}{\sqrt{\Delta T}} \sum_{n=N-N_1}^{N-1} e^{j2\pi f[N-1-n]\Delta W t_n} \end{cases} \tag{42}$$

The DF-PLFM chips comprised of two up-chirp segments in the presence of fading can be represented by,

$$u_f(t) = \begin{cases} \dfrac{\alpha_1}{\sqrt{\Delta T}} \sum_{n=0}^{\frac{N}{2}-1} e^{j2\pi f_1[n]\Delta W t_n} \\ \dfrac{\alpha_2}{\sqrt{\Delta T}} \sum_{n=\frac{N}{2}}^{N-1} e^{j2\pi f_2[N-1-n]\Delta W t_n} \end{cases} \tag{43}$$

$$d_f(t) = \begin{cases} \dfrac{\alpha_2}{\sqrt{\Delta T}} \sum_{n=0}^{\frac{N}{2}-1} e^{j2\pi f_2[n]\Delta W t_n} \\ \dfrac{\alpha_1}{\sqrt{\Delta T}} \sum_{n=\frac{N}{2}}^{N-1} e^{j2\pi f_1[N-1-n]\Delta W t_n} \end{cases} \tag{44}$$

The DF-PLFM chips comprised of an up-chirp and a down-chirp segment in the presence of fading can be represented by,

$$u_f(t) = \begin{cases} \dfrac{\alpha_1}{\sqrt{\Delta T}} \sum_{n=0}^{\frac{N}{2}-1} e^{j2\pi f_1[n]\Delta W t_n} \\ \dfrac{\alpha_2}{\sqrt{\Delta T}} \sum_{n=\frac{N}{2}}^{N-1} e^{j2\pi f_2[n]\Delta W t_n} \end{cases} \tag{45}$$

$$d_f(t) = \begin{cases} \dfrac{\alpha_2}{\sqrt{\Delta T}} \displaystyle\sum_{n=0}^{\frac{N}{2}-1} e^{j2\pi f_2\left[\frac{N}{2}-1-n\right]\Delta W t_n} \\ \dfrac{\alpha_1}{\sqrt{\Delta T}} \displaystyle\sum_{n=\frac{N}{2}}^{N-1} e^{j2\pi f_1[N-1-n]\Delta W t_n} \end{cases} \quad (46)$$

For the DF-LFM waveforms, all the discrete frequencies $0 \leq n \leq N_1 - 1$, are considered as the lower frequency component which are attenuated by $\alpha_1$, while the discrete frequencies $N_1 \leq n \leq N-1$ will be considered as the higher frequency component which are attenuated by $\alpha_2$. For the DF-PLFM waveforms, all the discrete frequencies $0 \leq n \leq \frac{N}{2} - 1$ are considered as the lower frequency component and $\frac{N}{2} \leq n \leq N - 1$, the higher frequency component. Figure 10 shows the autocorrelations of the DF-LFM chips with and without fading, i.e., $u_f(t)$ and $d_f(t)$ and $u(t)$ and $d(t)$ for $N = 16$, $N_1 = 12$, $N_g = 16$, $\alpha_1 = 0.9792$ and $\alpha_2 = 0.8470$ ($\alpha_1$ and $\alpha_2$ are randomly generated). Although, fading results in reducing the mainlobe peaks of the autocorrelations, $|R_{u_f}(t)|$ and $|R_{d_f}(t)|$ (as seen in the plots), they remain identical.

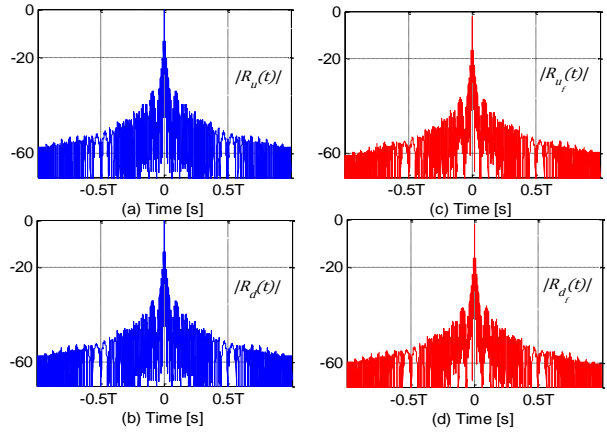

**Figure 10.** (**a**,**b**) Plots in dB of $|R_u(t)|$ and $|R_d(t)|$; and (**c**,**d**) plots of $|R_{u_f}(t)|$ and $|R_{d_f}(t)|$.

Figure 11 shows the matched filter outputs, $R_{x',a}(t)$ and $R_{x',b}(t)$, and their sum $R(t)$ (given by Equations (30), (33) and (37), respectively), in the presence of frequency selective fading. Clearly, the zero-sidelobe region on either side of the mainlobe is not affected by fading.

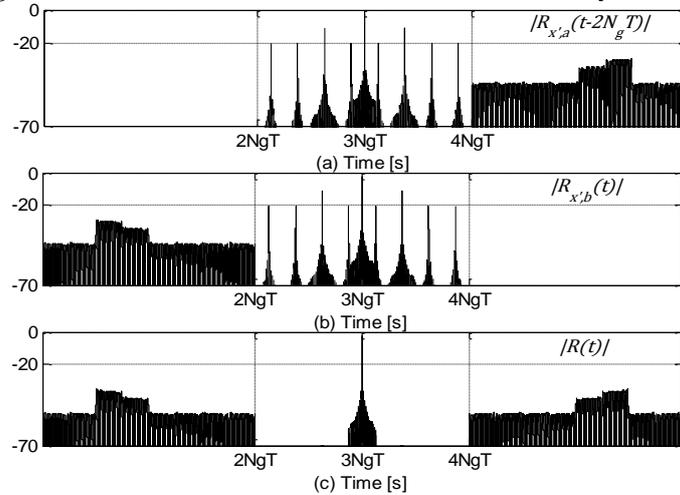

**Figure 11.** Plots in dB of: (**a**) $\dfrac{R_{x',a}(t-2N_gT)}{N_g}$; (**b**) $\dfrac{R_{x',b}(t)}{N_g}$; and (**c**) $\dfrac{|R(t)|}{2N_g}$, in the presence of frequency selective fading.

Root mean square error ($RMSE$) in the location of the mainlobe is used as a metric to measure of effectiveness in sidelobe cancellation in the presence of frequency selective fading and noise. Location of the mainlobe of $R(t)$ is given by,

$$\hat{t} = \underset{t}{\operatorname{argmax}} \left| \frac{R(t)}{2N_g} \right| \quad (47)$$

where $R(t)$ is the final output, given by Equation (37).

$RMSE$ in the location of the mainlobe can be represented by,

$$RMSE = \frac{1}{T}\left(\frac{1}{K}\sum_{i=1}^{K}(\hat{t}_i - t)^2\right)^{0.5} \quad (48)$$

where $K$ is the number of trials, $t$ is the actual timing of the mainlobe, $\hat{t}_i$ is the estimated timing of the mainlobe for the *ith* trial in the presence of noise and $T$ is the chip duration.

$RMSE$ vs. signal-to-noise ratio (SNR) plots for the proposed codes are shown in Figure 12.

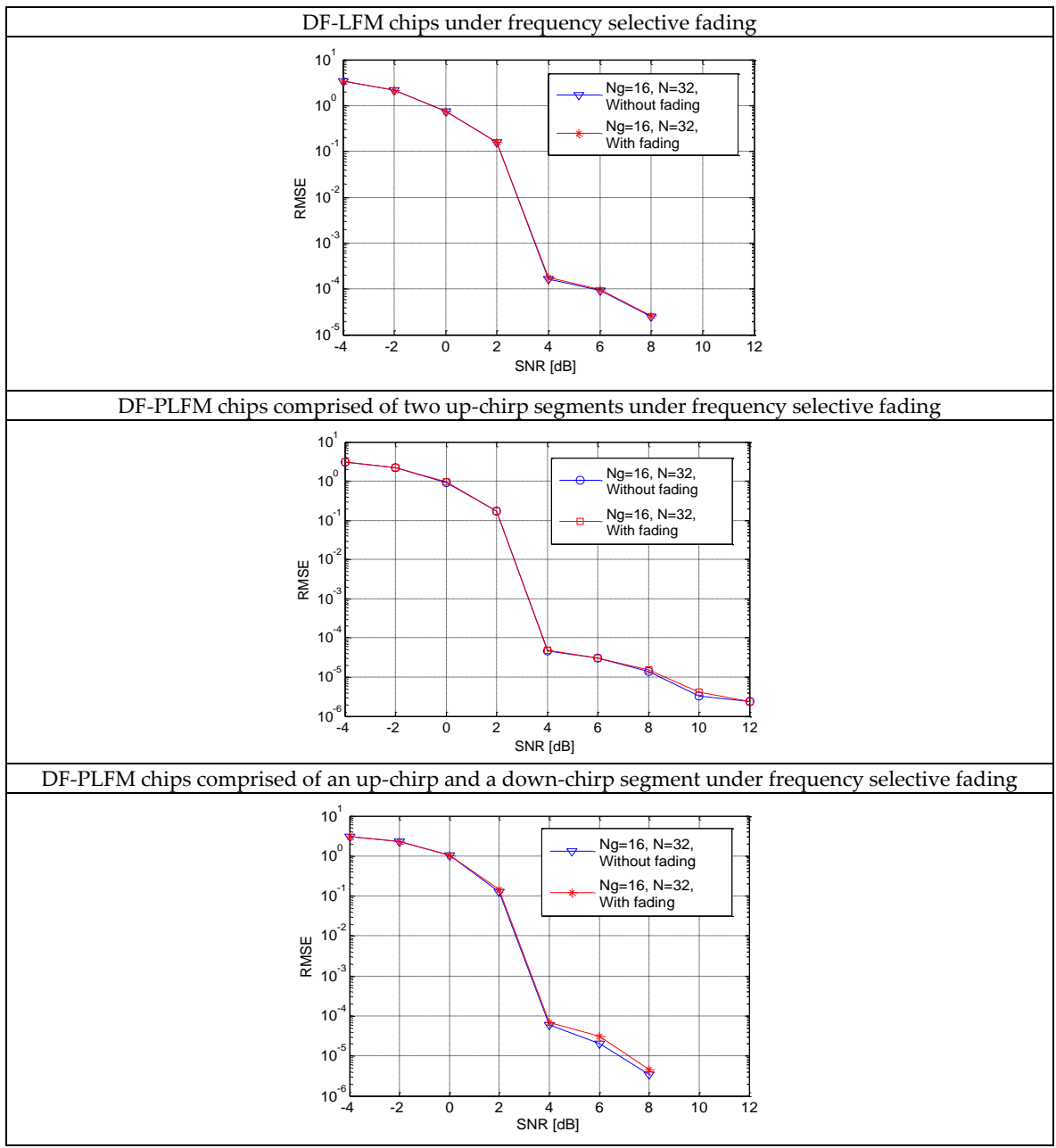

**Figure 12.** Plots of root mean square error ($RMSE$) vs. SNR for the proposed codes.

SNR is given by $20\log_{10}\frac{2N_g}{\sigma^2}$, where $2N_g$ is the code energy and $\sigma^2$ represents the noise power. SNR is varied by changing $\sigma^2$ and $K = 1000$ iterations are performed for each SNR value. $\alpha_1$ and $\alpha_2$ are randomly chosen such that $0.6 \leq \alpha_1, \alpha_2 \leq 1$. From these plots it can be observed that the RMSE curve with fading stays very close to the $RMSE$ in the absence of fading. This is true for the DF-LFM and the DF-PLFM chips.

As discussed in the Introduction, complete sidelobe cancellation can be achieved using a complementary code pair. However, it requires transmission and reception of the complementary code pair at two frequencies/channels as shown in Figure 1. In the presence of frequency selective fading, the cancellation of sidelobes is inexact which results in increasing the probability of false alarm. This can be seen in the plots of $RMSE$ vs. SNR in Figure 13 for a Golay complementary code pair $(A, B)$ of length $N_g = 16$ and 64. $A$ is attenuated by $\alpha_1$ and $B$ is attenuated by $\alpha_2$ such that $0.6 \leq \alpha_1, \alpha_2 \leq 1$. It can be observed that frequency selective fading results in increasing the RMSE in the location of the mainlobe. The proposed codes are resistant to the effect of frequency selective fading, as shown in the $RMSE$ plots in Figure 12.

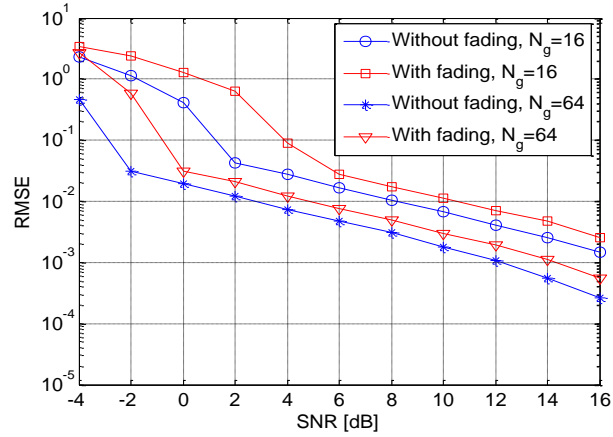

**Figure 13.** Plot of root mean square error ($RMSE$) vs. signal-to-noise ratio (SNR) for complementary code pair using sinusoidal chip.

## 6. Good Code Sets from Complementary Pair Using DF-LFM and DF-PLFM Waveforms as Chips

A good code set is constructed by changing the slopes of the instantaneous frequency (shown in the plots in Figure 3) of the DF-LFM/DF-PLFM waveforms used as chips for the same complementary code pair. The parameter $\gamma_l$, as shown in the Equations (20)-(25), is varied carefully to result in DF-LFM and DF-PLFM chips with different slopes. The resulting good code set has a zero domain on either side of the mainlobe and an adjacent region of small sidelobes. The cross-correlation between any code pair in the set is very small. The *lth* code in the set can be represented by,

$$s_l(t) = a_l(t) + b_l(t - 2N_g T) \qquad (49)$$

where $a_l(t) = \sum_{n=1}^{N_g} a[n]u_l(t - (n-1)T)$, $b_l(t) = \sum_{n=1}^{N_g} b[n]d_l(t - (n-1)T)$, and $u_l(t)$ and $d_l(t)$ represent the two symmetrical DF-LFM/DF-PLFM waveforms.

$u_l(t)$ and $d_l(t)$ for the good code sets from complementary pairs using DF-LFM chips can be represented by,

$$u_l(t) = \frac{1}{\sqrt{\Delta T}} \sum_{n=0}^{N-1} e^{j2\pi f_l[n]\Delta W t_n} \qquad (50)$$

$$d_l(t) = \frac{1}{\sqrt{\Delta T}} \sum_{n=0}^{N-1} e^{j2\pi f_l[N_l-1-n]\Delta W t_n} \qquad (51)$$

where $f_l[n] = round\left(\frac{2\gamma_l(N-1)n}{N}\right)$ for $n = 0, 1, 2, ..., N - 1$, $\gamma_l = k + \frac{(l-1)(1-3k)}{L-1}$, $l = 0, 1, 2, ..., L - 1$, $k \in (0, 0.5]$, $\Delta W = \frac{1}{\Delta T}$ and $n\Delta T \leq t_n \leq (n+1)\Delta T$.

$u_l(t)$ and $d_l(t)$ for the good code sets from complementary pairs using DF-PLFM chips comprised of two up-chirps can be represented by,

$$u_l(t) = \begin{cases} \frac{1}{\sqrt{\Delta T}} \sum_{n=0}^{\frac{N}{2}-1} e^{j2\pi f_{l,1}[n]\Delta W t_n} \\ \frac{1}{\sqrt{\Delta T}} \sum_{n=\frac{N}{2}}^{N-1} e^{j2\pi f_{l,2}[N-1-n]\Delta W t_n} \end{cases} \quad (52)$$

$$d_l(t) = \begin{cases} \frac{1}{\sqrt{\Delta T}} \sum_{n=0}^{\frac{N}{2}-1} e^{j2\pi f_{l,2}[n]\Delta W t_n} \\ \frac{1}{\sqrt{\Delta T}} \sum_{n=\frac{N}{2}}^{N-1} e^{j2\pi f_{l,1}[N-1-n]\Delta W t_n} \end{cases} \quad (53)$$

where $f_{l,1}[n] = round\left(\frac{2\gamma_l(N-1)n}{N}\right)$ for $n = 0, 1, 2, ..., \frac{N}{2} - 1$, $\gamma_l = k + \frac{(l-1)(1-2k)}{L-1}$, $f_{l,2}[n] = round\left((N-1)(1 - \frac{(1-\gamma_l)2n}{N-1})\right)$ for $n = \frac{N}{2}, \frac{N}{2} + 1, ..., N - 1$ and $k \in (0, 0.5]$.

$u_l(t)$ and $d_l(t)$ for the good code sets from complementary pairs using DF-PLFM chips comprised of an up-chirp and a down-chirp are given by,

$$u_l(t) = \begin{cases} \frac{1}{\sqrt{\Delta T}} \sum_{n=0}^{\frac{N}{2}-1} e^{j2\pi f_{l,1}[n]\Delta W t_n} \\ \frac{1}{\sqrt{\Delta T}} \sum_{n=\frac{N}{2}}^{N-1} e^{j2\pi f_{l,2}[n]\Delta W t_n} \end{cases} \quad (54)$$

$$d_l(t) = \begin{cases} \frac{1}{\sqrt{\Delta T}} \sum_{n=0}^{\frac{N}{2}-1} e^{j2\pi f_{l,2}[N_l-1-n]\Delta W t_n} \\ \frac{1}{\sqrt{\Delta T}} \sum_{n=\frac{N}{2}}^{N-1} e^{j2\pi f_{l,1}[N-1-n]\Delta W t_n} \end{cases} \quad (55)$$

The receiver for each code is as shown in Figure 6. The ACFs of all the codes in the set are as shown in the plots in Figure 7. Thus, the autocorrelation sidelobe peak ($ASP_l$) of the $lth$ code in the set is given by,

$$ASP_l = CCP_{a_l,b_l} \quad (56)$$

where $CCP_{a_l,b_l}$ is the same as $CCP_{a,b}$ which is defined in Equation (38) and shown in the plots in Figure 7.

Maximum cross-correlation peak ($MCCP$) and average cross-correlation peak ($ACCP$) are used as measures to compare the different good code sets introduced in this paper.

$$MCCP = \max_{l,k=1,2,...,L} \left\{ \max_t |R_{s_l,s_k}(t)| \right\} \quad (57)$$

$$ACCP = \frac{\sum_{l=1}^{L} \sum_{k=l}^{L} \max_t |R_{s_l,s_k}(t)|}{\binom{L}{2}} \quad (58)$$

where $R_{s_l,s_k}(t)$ is the normalized cross-correlation between the code pair $\{l, k\}$ in the set.

The table (Column 2) in Figure 14 shows the plots of $MCCP$ and $ACCP$ for each set for different values of $N$, $N_g$ and $L$.

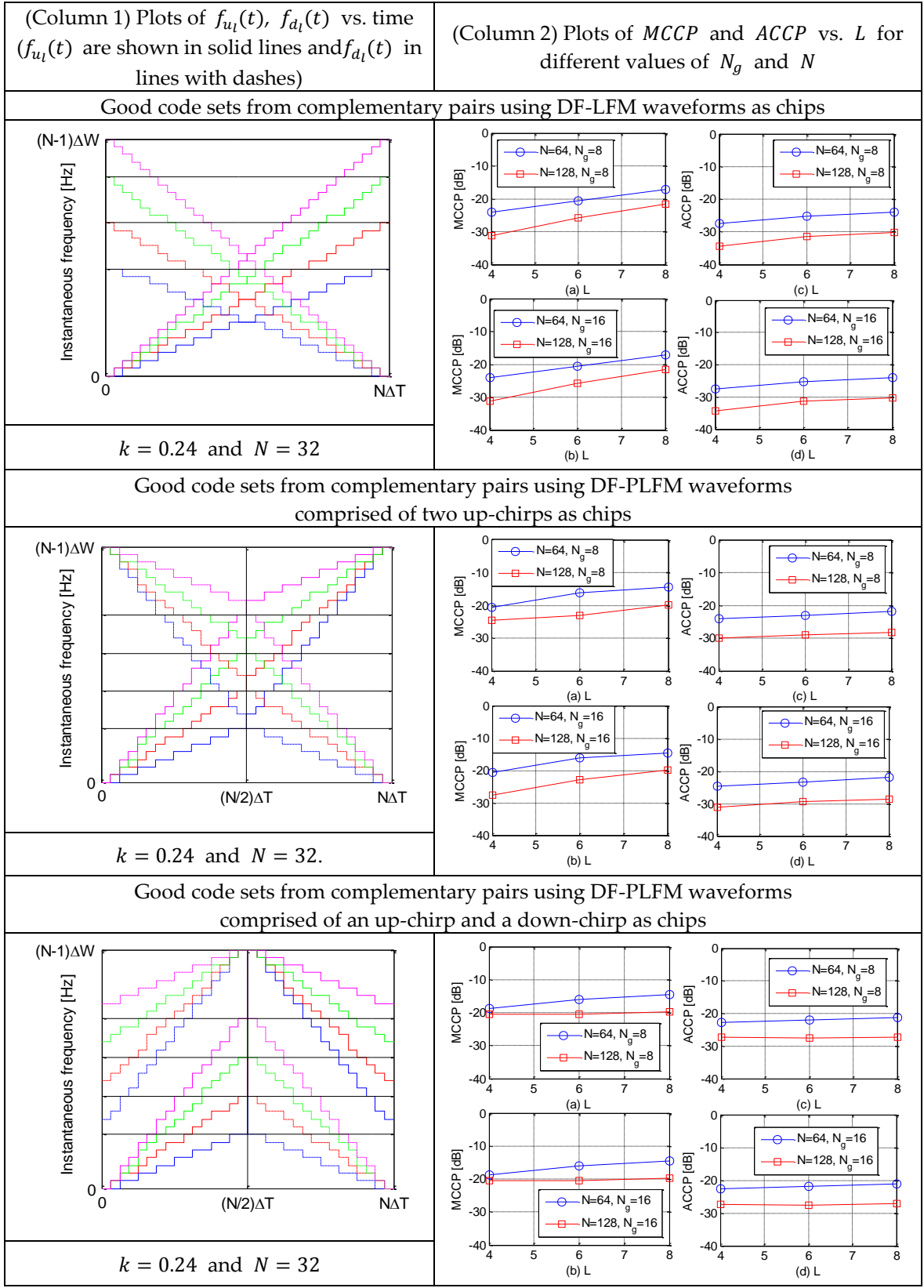

**Figure 14.** Plots of instantaneous frequencies, i.e., $f_{u_l}(t)$ and $f_{d_l}(t)$ vs. time in column 1; and (a), (b) plots of maximum cross-correlation peaks ($MCCP$) and (c), (d) average cross-correlation peaks ($ACCP$) vs. number of codes ($L$), for various values of code length ($N_g$) and number of discrete frequencies ($N$) in column 2.

From the plots in Figure 14 it can be observed that the good code sets constructed using DF-LFM chips have better $MCCP$ and $ACCP$ properties compared to those constructed using

DF-PLFM chips. However, the time bandwidth products of all the codes in the sets constructed using DF-PLFM chips is the same (i.e., $N(N-1)\Delta W \Delta T$), while those of the codes in the set constructed using DF-LFM chips are different. In addition, for a given complementary code pair code length ($N_g$), the cross-correlation sidelobe peaks could be reduced by increasing the number of discrete frequencies ($N$) in the chips.

## 7. Construction of a Good Code Set from the Mates of a Complementary Code Pair

Consider the complementary code pair: $B(z^{-1})$ and $-A(z^{-1})$. The sum of the cross-correlation of $A(z)$ with $B(z^{-1})$ and that of $B(z)$ with $-A(z^{-1})$ will be,

$$A(z)B(z) - B(z)A(z) = 0 \tag{59}$$

Thus, the code pair: $B(z^{-1})$ and $-A(z^{-1})$, are orthogonal to the complementary code pair: $A(z)$ and $B(z)$. Such code pairs are called complementary code pair mates, as described in [23,24].

Let $S_1(z) = A(z^{N_c})C(z) + z^{-D}B(z^{N_c})\gamma C(z^{-1})$ and $S_2(z) = B(z^{-N_c})C(z) - z^{-D}A(z^{-N_c})\gamma C(z^{-1})$ be the codes constructed by concatenating the complementary code pair mates ($A(z), B(z)$ and $B(z^{-1}), -A(z^{-1})$) using $C(z)$ and its symmetrical/anti-symmetrical mirror image ($\gamma C(z^{-1})$, where $\gamma = \pm 1$) as chips, as described in Section 2. Their autocorrelations can be represented by,

$$R_{S_1}(z) = \gamma R_{A,B}(z^{N_c})C^2(z) + 2N_g z^{-D} R_c(z) + \gamma z^{-2D} R_{B,A}(z^{N_c})C^2(z^{-1}) \tag{60}$$

$$R_{S_2}(z) = \gamma R_{A,B}(z^{-N_c})C^2(z) + 2N_g z^{-D} R_c(z) - \gamma z^{-2D} R_{B,A}(z^{-N_c})C^2(z^{-1}) \tag{61}$$

Both the autocorrelations consist of a zero-sidelobe region on either side of the mainlobe and an adjacent region of small cross-correlation sidelobes, as shown in Equation (37) in Section 2. Their autocorrelation sidelobe peaks (i.e., $CCP$, given by (38)) vary with $N$ and $N_g$ as shown in the plots in Figure 7. The cross-correlation between $S_1(z)$ and $S_2(z)$ can be represented by,

$$R_{S_1,S_2}(z) = (A(z^{N_c})C(z) + z^{-D}B(z^{N_c})\gamma C(z^{-1}))(B(z^{N_c})C(z^{-1}) - z^D A(z^{N_c})\gamma C(z^1)) \tag{62}$$

Simplifying Equation (62) results in,

$$R_{S_1,S_2}(z) = -A^2(z^{N_c})C^2(z) + 0 + \gamma z^{-2D} B^2(z^{N_c})C^2(z^{-1}) \tag{63}$$

Clearly, the cross-correlation consists of a zero-sidelobe region and an adjacent region of small cross-correlation sidelobes. Thus, $S_1(z)$ and $S_2(z)$ are quazi-orthogonal in the convolution sense. The zero-sidelobe domain in the cross-correlation is due to the symmetry/anti-symmetry property of the chips. Since, the symmetry/anti-symmetry property is immune to frequency selective fading, the zero-sidelobe domain in the cross-correlation is also immune to frequency selective fading. Figure 15 show the autocorrelations $S_1(z)$ and $S_2(z)$ and their cross-correlation. For these plots, $S_1(z)$ and $S_2(z)$ are constructed using DF-LFM chips with $N = 8$ and $N_g = 16$. These plots clearly show the zero-sidelobe domain in the cross-correlation between the proposed code constructed using a complementary code pair and the one constructed using its mate.

Varying the slopes of the DFCW chips in both $S_1(z)$ and $S_2(z)$, as shown in Section 6, results in two good code sets. Since these two sets are quasi-orthogonal to each other, they could be combined to form a larger set with twice the number of codes, while maintaining the same cross-correlation properties as those of the individual sets. This can be seen in the plots of maximum cross-correlation peaks ($MCCP$) and average cross-correlation peaks ($ACCP$) in Column 1 of the table in Figures 16 and 17. When compared to the plots in Figure 14, clearly the cross-correlation properties remain the same, while doubling the number of the codes in the set.

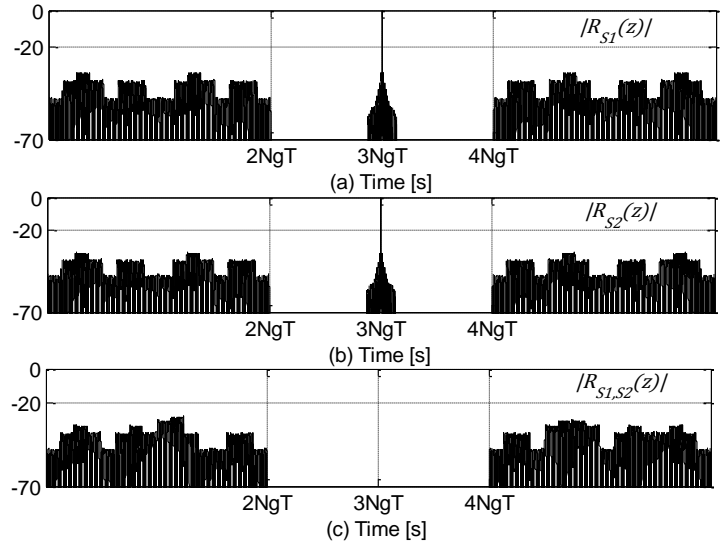

**Figure 15.** Plots in dB of: (**a**) $R_{S_1}(z)$; (**b**) $R_{S_2}(z)$; and (**c**) $R_{S_1,S_2}(z)$, for $N = 8$ and $N_g = 16$ using DF-LFM chips.

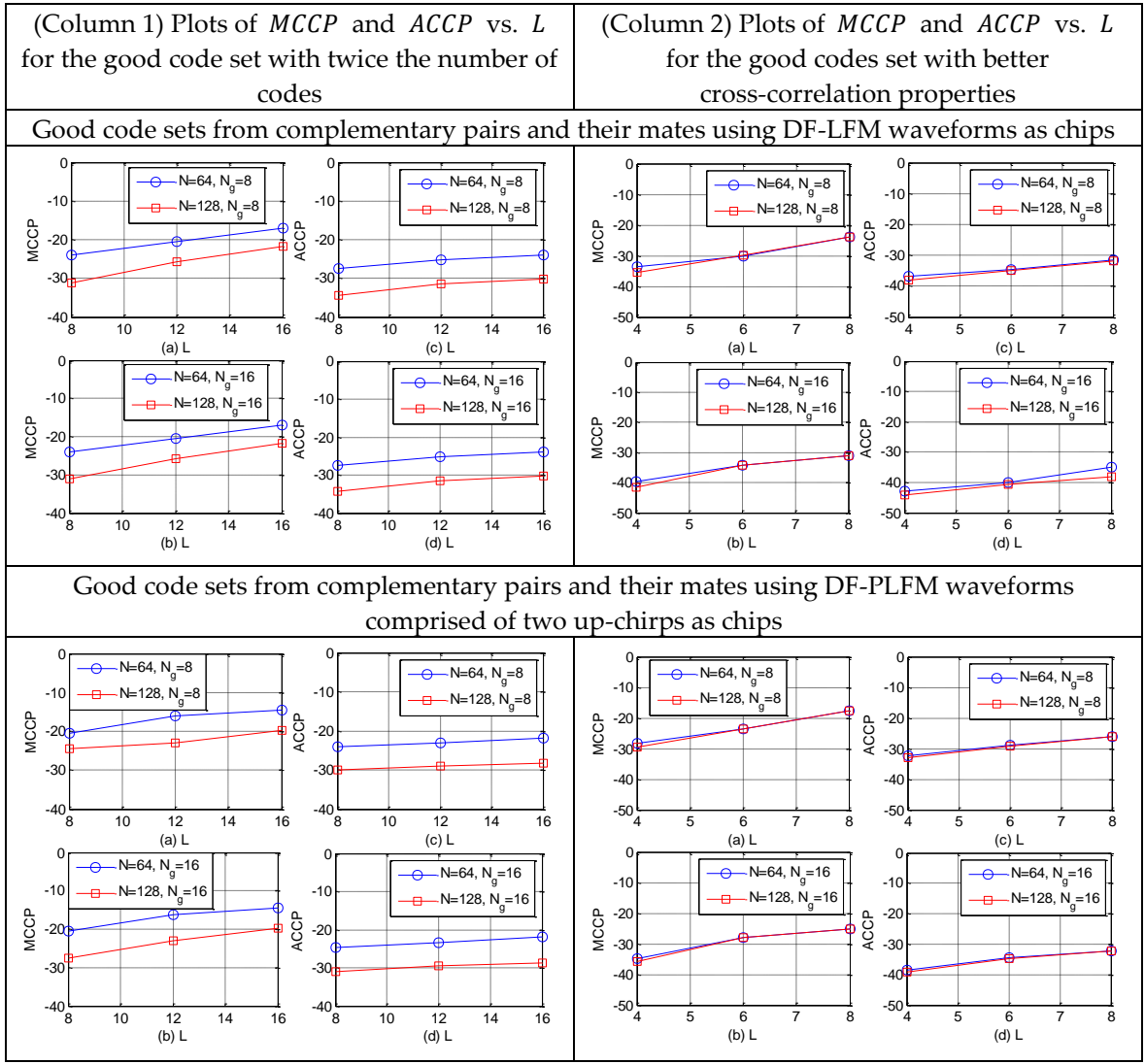

**Figure 16.** Plots of (a), (b) $MCCP$ vs. $L$ and (c), (d) $ACCP$ vs. $L$.

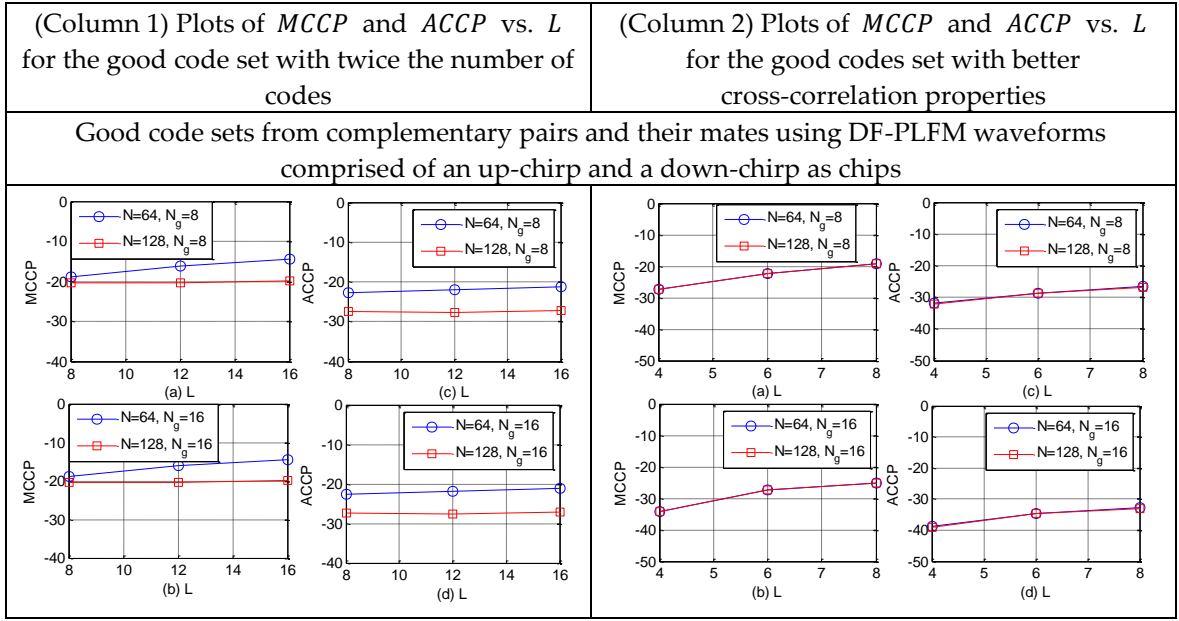

**Figure 17.** Plots of (a), (b) $MCCP$ vs. $L$ and (c), (d) $ACCP$ vs. $L$.

If a better good code set with the smaller $MCCP$ and $ACCP$ is required, then the best candidates from the two sets could be selected. This could be achieved by using only half the slope options for the DFCW waveforms used as chips for the complementary code pair and its mates i.e., the DF-LFM/DF-PLFM waveforms with $l = 0, 2, \ldots, L-1$ in Equation (52)–(57). This results in a good code set with significantly better $MCCP$ and $ACCP$ values, as shown in the plots in column 2 of the table in Figures 16 and 17. These are significantly smaller than the $MCCP$ and the $ACCP$ values in Figure 14 for the same number of codes in the set. Thus, constructing the proposed codes from complementary code pair and their mates, while using the same chips for both pairs, offers the aforementioned flexibility of choosing between a better good code set and a larger good code set.

Table 1 shows the average autocorrelation and average cross-correlation values for the Deng's polyphase [28] good code set and the Deng's discrete frequency [20] set designs. The autocorrelation and cross-correlation sidelobe peaks of the codes introduced in this paper are smaller, as shown in the plots in Figures 7 and 17 respectively. The proposed codes constructed via concatenating biphase Golay complementary code pairs and using discrete frequency chips results in a zero-sidelobe domain and an adjacent region of very small cross-correlation sidelobes. In [28], it is shown that the average autocorrelation sidelobe and cross-correlation peaks decrease with increase in code length and approach $O(\frac{1}{\sqrt{N}})$ for large $N$. Their discrete frequency good code set [20] being wideband, achieve much smaller sidelobe peaks. In the case of the good code set design proposed in this paper, for a given code length ($N_g$), increasing the number of discrete frequencies ($N$) in the chips results in a better code set.

| Good code set | Average autocorrelation peak (dB) | Average cross-correlation peak (dB) |
|---|---|---|
| Deng's polyphase set (code length = 128 and $L = 3$ codes) | −20.9606 | −19.1525 |
| Deng's discrete frequency set (# of discrete frequencies = 128 and $L = 3$ codes) | −32.2641 | −32.2522 |
| Proposed code set with $N_g = 16$, $N = 32$ and $L = 4$ | −39.8280 | −37.9926 |

**Table 1.** Table containing the average autocorrelation and cross-correlation peaks for the proposed set in comparison with Deng's polyphase and discrete frequency code sets.

## 8. Conclusions

In this paper, it is shown that, by replacing the sinusoidal chips in the complementary code pair with waveforms that satisfy two conditions, symmetry/anti-symmetry and quazi-orthogonality in the convolution sense, allows for concatenating them in the time domain using one frequency band/channel. This results in a zero sidelobe domain around the mainlobe and an adjacent region of small cross-correlation sidelobes. Symmetry/anti-symmetry property results in the zero-sidelobe region, while the quasi-orthogonality property makes the adjacent region of cross-correlation sidelobes small. Discrete frequency coding waveform (DFCW) based on LFM and PLFM waveforms are used as chips, since they satisfy both the symmetry/anti-symmetry and quasi-orthogonality conditions. Since, frequency selective fading does not affect the symmetry/anti-symmetry property, the zero-sidelobe region is resistant to the effects of fading. A good code set with a zero-sidelobe region is then constructed by varying the slopes of the DFCW chips, while using the same complementary code pair. Such good code sets are constructed using a set of quasi-orthogonal DFCW chips based on LFM and PLFM waveforms. It is also shown that mates of the complementary code pair could be used to generate a second good code set using the same DFCW chips. These two sets are shown to be quasi-orthogonal with a zero-sidelobe domain. Thus, resulting in a larger good code set with twice the number of codes, while maintaining the same cross-correlation properties. Or a better good code set could be constructed by choosing the best candidates from the two sets.

As a topic for further research, the proposed codes could be constructed using polyphase versions (as shown in [29,30]) of the LFM/PLFM waveforms as chips with the complementary code pairs and their mates. Exploring other waveforms or codes that satisfy the two conditions of symmetry/anti-symmetry and quazi-orthogonality and improving the Doppler properties of these codes are the other topics of future research.

## References

1. Frank, R.C. Polyphase codes with good nonperiodic correlation properties. *IEEE Trans. Inf. Theory* **1963**, *9*, 43–45.
2. Costas; J.P. A study of a class of detection waveforms having nearly ideal range-Doppler ambiguity properties. *Proc. IEEE* **1984**, *72*, 996–1009.
3. Golomb, S.W.; Taylor, H. Construction and properties of Costas arrays. *Proc. IEEE* **1984**, *72*, 1143–1163.
4. Barker, L.; Drakakis, K.; Rickard, S. On the Complexity of the Verification of the Costas Property. *Proc. IEEE* **2009**, *97*, 586–593.
5. Fam, A.T.; Sarkar, I.; Poonnen, T. Area and power efficient mismatched filters based on sidelobe inversion. In Proceedings of the IEEE 2008 Radar Conference, Rome, Italy, 26–30 May 2008; pp. 1–6.
6. Akbaripour, A; Bastani, M.H. Range Sidelobe Reduction Filter Design for Binary Coded Pulse Compression System. *IEEE Trans. Aerosp. Electron. Syst.* **2012**, *48*, 348–359.
7. Jung, K.T.; Kim, C.J.; Lim, C.H.; Lee, H.S.; Kwag, Y.K. Design of optimum mean square sidelobe suppression filters for Barker codes. In Proceedings of the 1992 International Conference on Radar, Brighton, UK, 12–13 October 1992; pp. 530–533.
8. Rihaczek, A.W.; Golden, R.M. Range Sidelobe Suppression for Barker Codes. *IEEE Trans. Aerosp. Electron. Syst.* **1971**, *AES-7*, 1087–1092.
9. Fam, A.T.; Qazi, F.A; Kadlimatti, R. Zero Sidelobe Aperiodic Codes via Additive-Multiplicative Mismatched Filtering. In Proceedings of the 2013 IEEE Military Communications Conference (MILCOM 2013), San Diego, CA, USA, 18–20 November 2013; pp. 829–836.
10. Golay, M.J.E. Complementary series. *IRE Trans. Inf. Theory* **1961**, *7*, 82–87.
11. Frank, R. Polyphase complementary codes. *IEEE Trans. Inf. Theory* **1980**, *26*, 641–647.
12. Jafari, A.H.; O'Farrell, T. Performance Evaluation of Spatial Complementary Code Keying Modulation in MIMO Systems. In Proceedings of the 2015 IEEE 81st Vehicular Technology Conference (VTC Spring), Glasgow, UK, 11–14 May 2015; pp. 1–5.

13. Galati, G.; Pavan, G. Range Sidelobes Suppression in Pulse-Compression Radar using Golay Pairs: Some Basic Limitations for Complex Targets. *IEEE Trans. Aerosp. Electron. Syst.* **2012**, *48*, 2756–2760.
14. Liu, Z.; Parampalli, U.; Guan, Y.L. Optimal Odd-Length Binary Z-Complementary Pairs. *IEEE Trans. Inf. Theory* **2014**, *60*, 5768–5781.
15. Tang, X.; Mow, W.H. Design of spreading codes for quasi-synchronous CDMA with intercell interference. *IEEE J. Sel. Areas Commun.* **2006**, *24*, 84–93.
16. Li, D. The perspectives of large area synchronous CDMA technology for the fourth-generation mobile radio. *IEEE Commun. Mag.* **2003**, *41*, 114–118.
17. Qazi, F.A.; Fam, A.T. Good code sets based on Piecewise Linear FM. In Proceedings of the 2012 IEEE Radar Conference, Atlanta, GA, USA, 7–11 May 2012; pp. 522–527.
18. Fishler, E.; Haimovich, A.; Blum, R.; Chizhik, D.; Cimini, L.; Valenzuela, R. MIMO radar: An idea whose time has come. In Proceedings of the 2004 IEEE Radar Conference, Philadelphia, PA, USA, 26–29 April 2004; pp. 71–78.
19. Deng, H. Orthogonal netted radar systems. *IEEE Aerosp. Electron. Syst. Mag.* **2012**, *27*, 28–35.
20. Deng, H. Discrete frequency-coding waveform design for netted radar systems. *IEEE Signal Proc. Lett.* **2004**, *11*, 179–182.
21. Qazi, F.A.; Fam, A.T. Discrete Frequency-Coding Waveform sets based on Piecewise Linear FM. In Proceedings of the 2014 IEEE Radar Conference, Cincinnati, OH, USA, 19–23 May 2014; pp. 469–473.
22. Liu, B.; He, Z.; He, Q. Optimization of Orthogonal Discrete Frequency-Coding Waveform Based on Modified Genetic Algorithm for MIMO Radar. In Proceedings of the 2007 International Conference of Communications, Circuits and Systems (ICCCAS), Kokura, Japan, 11–13 July 2007; pp. 966–970.
23. Harris. F.; Dick. C. A versatile Filter Structure to Generate and Compress Binary and Polyphase Complementary Spreading Codes. In Proceedings of the 2004 SDR Technical Conference and Product Exposition, Phoenix, AZ, USA, 15–18 November 2004.
24. Suehiro, N. Complete complementary code composed of N-multiple-shift orthogonal sequences. *IECE Trans.* **1982**, *J65 A*, 1247–1253.
25. Fam, A.T.; Kadlimatti, R. Complementary code pairs that share the same bandwidth via symmetrical linear FM chips. In Proceedings of the 2015 IEEE Military Communications Conference (MILCOM), Tampa, FL, USA, 26–28 October 2015; pp.820–825.
26. Kadlimatti, R.; Fam, A.T. Good code sets from complementary pairs via symmetrical/anti-symmetrical chips. *IEEE Trans. Aerosp. Electron. Syst.* **2016**, *52*, 1327–1339.
27. Pezeshki, A.; Calderbank, A.R.; Moran, W.; Howard, S.D. Doppler Resilient Golay Complementary Waveforms. *IEEE Trans. Inf. Theory* **2008**, *54*, 4254–4266.
28. Deng, H. Polyphase code design for Orthogonal Netted Radar systems. *IEEE Trans. Signal Proc.* **2004**, *52*, 3126–3135.
29. Lewis, B.L.; Kretschmer, F.F. Linear Frequency Modulation Derived Polyphase Pulse Compression Codes. *IEEE Trans. Aerosp. Electron. Syst.* **1982**, *AES-18*, 637–641.
30. Qazi, F.A.; Fam, A.T. Doppler tolerant and detection capable polyphase code sets. *IEEE Trans. Aerosp. Electron. Syst.* **2015**, *51*, 1123–1135.